\documentclass[prc,amsmath,twocolumn,showkeys,superscriptaddress]{revtex4}
\usepackage{gensymb}
\usepackage{graphicx,color}
\usepackage{amssymb}
\usepackage{enumerate}
\usepackage{verbatim}
\usepackage{natbib}
\usepackage{float}
\usepackage[outdir=./]{epstopdf}
\usepackage[dvipsnames]{xcolor}

\begin{document}
  \newcommand {\nc} {\newcommand}
  \nc {\Sec} [1] {Sec.~\ref{#1}}
  \nc {\IR} [1] {\textcolor{red}{#1}} 
  \nc {\IB} [1] {\textcolor{blue}{#1}}
  \nc {\IV} [1] {\textcolor{violet}{#1}} 
  \nc {\IG} [1] {\textcolor{green}{#1}}

\title{Properties of a separable representation of optical potentials}

\author{M. Quinonez}
\affiliation{National Superconducting Cyclotron Laboratory, Michigan State University, East Lansing, MI 48824}
\affiliation{Department of Physics and Astronomy, Michigan State University, East Lansing, MI 48824-1321}
\author{L. Hlophe}
\affiliation{National Superconducting Cyclotron Laboratory, Michigan State University, East Lansing, MI 48824}
\affiliation{Department of Physics and Astronomy, Michigan State University, East Lansing, MI 48824-1321}
\affiliation{Lawrence Livermore National Laboratory, L-414, Livermore, CA 94551}
\author{F.~M.~Nunes}
\email{nunes@nscl.msu.edu}
\affiliation{National Superconducting Cyclotron Laboratory, Michigan State University, East Lansing, MI 48824}
\affiliation{Department of Physics and Astronomy, Michigan State University, East Lansing, MI 48824-1321}

\date{\today}


\begin{abstract}
\begin{description}
\item[Background:] Separable interactions have a long history in nuclear physics. In the last few years, separable expansions have been used to represent the optical potential between a nucleon (proton or neutron) and a target.
\item[Purpose:] We explore the non-local properties of these separable optical potentials as well as their convergence behavior. 
\item[Method:] For a couple of cases, we use the generalized Ersnt-Shakin-Thaler scheme to generate separable interactions starting from local optical potentials. We study the variation of the interaction with energy range and rank. 
\item[Results:]  We find that, overall the off-diagonal behavior of the converged separable interaction deviates from the Gaussian form assumed by Perey and Buck. However, in the region surrounding the maximum depth the Gaussian form works quite well.  Focusing on this region, we study potentials describing neutron elastic scattering on $^{16}$O and $^{48}$Ca for beam energies in the range of $ E=$10-50~MeV and explore several measures of non-locality of the separable interactions.
 
\item[Conclusions:]  When the energy range considered for generating the separable interaction is $0\le E_{range}\le 50$ MeV, the resulting non-locality is large and target dependent.  Contrarily, the nonlocality obtained including larger energy ranges in the separable procedure is independent of the target and other details of the original local potential.
We find that, even when including in the expansion many support points with energy ranges $0\le E_{range}\le 2400$~MeV, the resulting potential retains non-local behavior.
Connections with microscopic optical potentials as well as other transformations used in the nucleon-nucleon domain are made.
\end{description}
\end{abstract}


\maketitle

\section{Introduction}
\label{intro}

One of the greatest challenges in the physics of nuclei concerns the interactions themselves. Effective interactions are developed to incapsulate the many degrees of freedom contained in the system. Much work has been devoted to the development of both nucleon-nucleon effective interactions \cite{nnforce}, the so-called NN force, and nucleon-nucleus effective interactions, referred to as optical potentials  (e.g. \cite{Hodgson_1984,ch89,kd2003}). In this work we focus on the latter,  and in particular on their properties when represented in separable form.

In the past, NN forces were derived phenomenologically with different levels of complexity (e.g. AV18 \cite{av18} and Minnesota \cite{minnesota}). In the last two decades the field has shifted toward generating these interactions in a more controlled fashion  through effective field theory (EFT) \cite{machleidt2020chiral}. Different transformations on NN forces have also been proposed to enable greater efficiency when used in many-body problems: these include $V_{low k}$ \cite{bogner2003,bogner2003plb} and Similarity Renormalization Group methods \cite{bogner2007}. In both of these examples high-momentum components of the interaction are shifted to low momentum off-diagonal behavior, while preserving the on-shell properties of the interaction. When analyzed in coordinate space, these transformations induce non-locality properties which do not affect the two-body observables but can have an impact in three- and more-body calculations. As will be discussed here, a similar situation can occur when considering nucleon-nucleus optical potentials. 

Separable interactions have a long history in few-nucleon physics (e.g. \cite{haidenbauer1984,koike1987,plessas2006}). Because the three-body Faddeev equations in momentum space \cite{ags1967} simplify greatly when using separable interactions, this approach was originally very popular. As computational capabilities increased, the few-nucleon field evolved to using more realistic non-separable interactions (e.g. \cite{fachruddin2000,witala2012}). The complications introduced by the infinite range Coulomb force in the three-nucleon problem were handled separately by screening and renormalization techniques \cite{deltuva2005}. 

For over a decade, the few-nucleon techniques have been ported into nuclear reactions and in particular to describe deuteron induced reactions \cite{deltuva2009}. Deuteron induced reactions are typically modeled as a three-body problem $n+p+A$, the input being the effective nucleon-target optical potentials. As was later realized, the Coulomb screening method introduced by the Lisbon group \cite{deltuva2005} could not be  applied to deuteron induced reactions involving heavy targets, due to the increased strength of the Coulomb force \cite{upadhyay2012}. It turns out that, by using a separable representation for the optical potential, those difficulties can be overcome \cite{akram2012}. As a result, in the last few years separable interactions have made a come back \cite{hlophe2013,hlophe2014,hlophe2017}.  These developments use the Ernst, Shakin, and Thaler  scheme (EST)  \cite{Ernst:1973zzb}  to generate separable representations for the nucleon optical potential. 

While most optical potentials being used to interpret nuclear reaction data are local \cite{ch89,kd2003}, separable interactions with realistic truncations are intrinsically non-local. Even though at the two-body level, the EST scheme ensures that scattering observables are exactly reproduced within a chosen energy range, this is not guaranteed when using these interactions in the context of deuteron-induced reactions, because the three-body equations will pick up off-shell contributions. It has been shown that local and non-local optical potentials can give rise to very different transfer cross sections even if they are equivalent at the two-body level \cite{ross2015,ross2016,titus2016,li2018}. It is therefore timely to perform a dedicated study on the effect that the separable EST transformation has on the properties of optical potentials. This is precisely the goal of this work. 

We study the scattering of neutrons on two closed shell nuclei $^{16}$O and $^{48}$Ca at beam energies of experimental interest, and explore the nonlocality properties of the separable interactions in coordinate space. This paper is organized in the following way. In Section II we briefly introduce the EST method and the standard Gaussian non-locality form used to extract the non-local parameter for the interaction. In Section III we present the results obtained for both targets and discuss these results in the context of previous work. Finally, the conclusions are presented in Section IV.

\section{Theoretical considerations}
\label{theory}

Deuteron induced reactions on intermediate to heavy mass targets $A$ are treated as three body problems consisting of $n+p+A$. In such cases, the three-body dynamics of the reaction is generated from the pairwise interactions: $V_{np}$ reproducing the properties of the deuteron and its continuum, and the nucleon optical potentials $U_{nA}$ and $U_{pA}$,  typically describing nucleon scattering from the target $A$. In general, these optical potentials are energy dependent and contain an important imaginary term that effectively takes into account the removal of flux from the incident channel into other channels in the reaction that are not explicitly included. 

Although there have been many efforts to derive the optical potential from first principles, the common practice is to use a larger set of elastic data to fit it \cite{ch89,kd2003}. For convenience these potentials are most often made local, although isolated studies have been performed to include nonlocality in these interactions \cite{perey-buck}. 
For simplicity, in this study, we focus on the neutron-target potentials, although the results can be trivially generalized to  proton-target potentials.


To construct separable representations of the $n-A$ optical potentials $U_{nA}$, the generalized EST scheme of~\cite{Hlophe:2015rqn}
is adopted.  Although the original EST scheme focused only on  Hermitian potentials, the generalization presented in ~\cite{Hlophe:2015rqn}
extends its applicability to potentials that are complex and energy dependent. Since this work focuses only on $n-A$ interactions, we shall refer to these as $U$ and drop the $nA$ subscript hereafter. 

The key features of the generalized EST separable expansion
can be summarized as follows. First, one defines the states $|\psi^{(+)}_{E_i}\rangle$ and $|\psi^{(-)}_{E_i}\rangle$ , which are eigenstates
of the Hamiltonians  $H=H_0+U$  and  $H^*=H_0+U^*$, respectively,  with  eigenvalues $E_i \ge 0$ and $H_0$ being the free Hamiltonian.  The states
 $|\psi^{(+)}_{E_i}\rangle$ are the usual scattering wavefunctions fulfilling outgoing boundary conditions, while the asymptotic behavior 
 of $|\psi^{(-)}_{E_i}\rangle$ is that of an incoming spherical wave.  Second, the two-body potential $U$ is expanded using the basis states 
 $\{|\psi^{(+)}_{E_i\alpha}\rangle\}$ and $\{|\psi^{(-)}_{E_i\alpha}\rangle\}$,  leading to the partial wave separable potential
\begin{equation}
u_{\alpha} ( E) = \sum\limits_{i,j=1}^{N} |h_{i\alpha }\rangle   \lambda_{ij}^{\alpha}(E) \langle \tilde h_{j\alpha}| \ , 
\label{eq:2a1}
\end{equation}
where  $|h_{i\alpha}\rangle  \equiv  U_{\alpha }(E_i)|\psi^{(+)}_{E_i\alpha }\rangle$,  $|\tilde h_{i\alpha}\rangle  \equiv  U_{\alpha}^*(E)|\psi^{(-)}_{E_i\alpha}\rangle$,
and  $E$ is the two-body center of mass (c.m.) energy. Here $\alpha\equiv\{l j\}$ denotes a single channel with $l$ being the the orbital angular momentum and $j=|l\pm1/2|$ the total angular momentum. The  number of basis states $N$ defines the rank of the separable potential and the energy eigenvalues $E_i$ are called EST support points.
We note that  $|h_{i\alpha }\rangle$ and $\langle \tilde h_{i\alpha}|$ are related to the half-shell
transition ($t$) matrix by
\begin{eqnarray}
|h_{i\alpha}\rangle &=&U_{\alpha }(E_i)|\psi^{(+)}_{E_i\alpha}\rangle =t_{\alpha} ( E_i) |p_i\rangle\\
\langle \tilde h_{i\alpha}|&=&\langle\psi^{(-)}_{E_i\alpha }|U_{\alpha }(E_i)=\langle p_i|t_{\alpha} ( E_i),
\label{eq:2a2}
\end{eqnarray}
where $p_i =\sqrt{2\mu E_i}$ is the on-shell momentum, with $\mu$ being the reduced mass.
The absolute square of the on-shell $t$ matrix elements relates directly to the cross section for elastic scattering. 
The  half-shell $t$ matrix elements are obtained in momentum space by solving the Lippmann-Schwinger (LS) equation
\begin{equation}
t_{\alpha} ( E_i) |p_i\rangle= U_{\alpha}(E_i)|p_i\rangle+U_{\alpha}(E_i)G_0(E_i)t_{\alpha}(E_i)|p_i\rangle.
\label{eq.2a3}
\end{equation}
Negative energy EST support points can also be included in the expansion, and in that case the bound state wavefunctions replace the incoming and outgoing
scattering states (see \cite{hlophe2017} for details).

Finally, one defines the coupling matrix $ \lambda_{ij}^{\alpha}(E)$ by imposing the constraint
\begin{eqnarray}
 \langle\psi^{(-)}_{E_i\alpha }|&U_{\alpha }&(E)|\psi^{(+)}_{E_j\alpha }\rangle =\langle\psi^{(-)}_{E_i\alpha }|u_{\alpha }(E)|\psi^{(+)}_{E_j\alpha }\rangle  \nonumber \\
 &=&  \sum\limits_{n,m=1}^{N}  \langle\psi^{(-)}_{E_i\alpha }| h_{n\alpha }\rangle   \lambda_{nm}^{\alpha}(E) \langle \tilde h_{m \alpha} | \psi^{(+)}_{E_j\alpha }\rangle. 
\label{eq:2a4}
\end{eqnarray}
This definition of  $ \lambda_{ij}^{\alpha}(E)$  ensures that the matrix elements of the original  potential $U(E)$ and the separable potential $u_{\alpha}(E)$ between the basis states
are identical for all energies $E$.  For the special case where $E$ corresponds to one of the EST support points, Eq.~(\ref{eq:2a4}) implies that the eigenstates of
$H_0+u(E_i)$ coincide with those of $H_0+U(E_i)$.  This guarantees that the wavefunctions obtained using the original potential $U$ are identical to the ones  computed with its separable representation $u$ at the EST support points. This  is a crucial property of the original EST scheme, and is by construction preserved in the generalized expansion for complex potentials.





\section{Results}
\label{results}

We consider the energy dependent CH89 global optical potential \cite{ch89} and apply the EST scheme to produce separable forms.
We analyze the properties of the resulting potential at two scattering energies, $E=5$ MeV and $E=20$ MeV, which span beam energies of experimental interest for applications involving transfer reactions. We consider both the number of support points included in the expansion (the rank $N$) and the energy range $E_{range}$ for the support points.  
Support points are chosen wisely based on the structure of the two-body continuum. One can consider that effectively in EST we are interpolating the S-matrix S(E) (or T-matrix) and, as such, we need to choose the set {$E_i$} that will enable the reproduction of the original S(E).

In Table \ref{tab:est}  the specifications of the EST parameters, including the rank used in the expansions and the energies corresponding to the support points, are provided.
When a range of energy is given, it means that an even spacing of support points within that range were included.
\begin{table}[b]
    \centering
    \begin{tabular}{|l |c|c|c|}
    \hline
     & $E_{range}$ [MeV] & N & Support points $E_{i}$ [MeV] \\
      \hline
    EST10-Ca & 10 & 10 & 0.5 MeV 				; 0.5-10 \\
    EST40-Ca & 40 & 12 & 0.5, 7 MeV 				; 7-40 \\
    EST400-Ca & 400 & 14 & 0.5, 10, 30, 60, 100; 100-400 \\
    EST800-Ca &  800 & 22 & 0.5, 10, 30, 60, 100; 100-800 \\
    EST1200-Ca &  1200 & 27 & 0.5, 10, 30, 60, 100; 100-1200 \\
    EST1600-Ca &  1600 & 35 & 0.5, 10, 30, 60, 100; 100-1600 \\
    EST2000-Ca &  2000 & 45 & 0.5, 10, 30, 60, 100; 100-2000 \\
    EST2400-Ca &  2400 & 53 & 0.5, 10, 30, 60, 100; 100-2400 \\   
      \hline    
     \end{tabular}
         \caption{The EST parameters for the n-$^{48}$Ca separable potentials. $E_{range}$ specifies the highest support point used, and $N$ is the number of EST support points needed for convergence with a given $E_{range}$. The specific energies of the support points are shown in the last column; when a range of energy is given, it means that an even spacing of support points within that range were included. }
    \label{tab:est}
\end{table}
While a rank $N<10$ is usually sufficient to describe  nucleon scattering observables  up to  20 MeV, a much higher rank is needed  to reach convergence  of the potential matrix elements $u_{\alpha}(r',r)$. Thus we performed calculations up to $N=53$.  We find that the separable interactions obtained depend strongly on the energy range included in the EST procedure. Again, to fully explore this dependence, we consider multiple values of $E_{range}$, going all the way up to $2400$ MeV.

\begin{figure}[t]
\begin{center}
\includegraphics[width=0.5\textwidth]{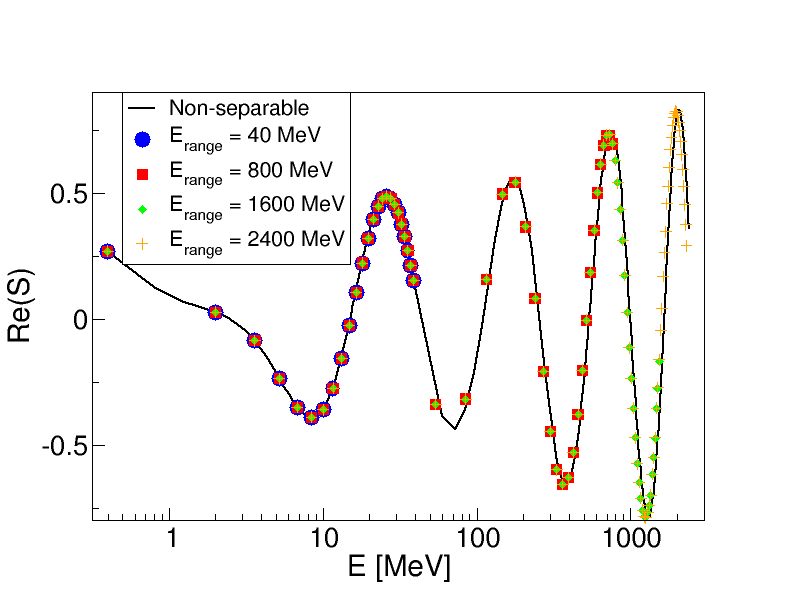}
\end{center}
\caption{[Color online] Real part of the S-matrix as a function of the scattering energy for the various energy ranges considered in the EST expansion (example shown for $^{48}$Ca(n,n) scattering $\ell =0$.}
 \label{fig-smat-ca}
\end{figure}

The S-matrices generated with the separable interactions  agree with those obtained directly with the original CH89 potential within their corresponding energy range but expectedly fail to provide an accurate description outside their energy range. Plotted in Fig. \ref{fig-smat-ca} is the real part of the S-matrix resulting from separable interactions with different $E_{range}$, for s-wave neutrons scattering off $^{48}$Ca. 

\subsection{Radial dependence of the separable interaction}

\begin{figure}[t!]
\begin{center}
    \includegraphics[width=0.45 \textwidth]{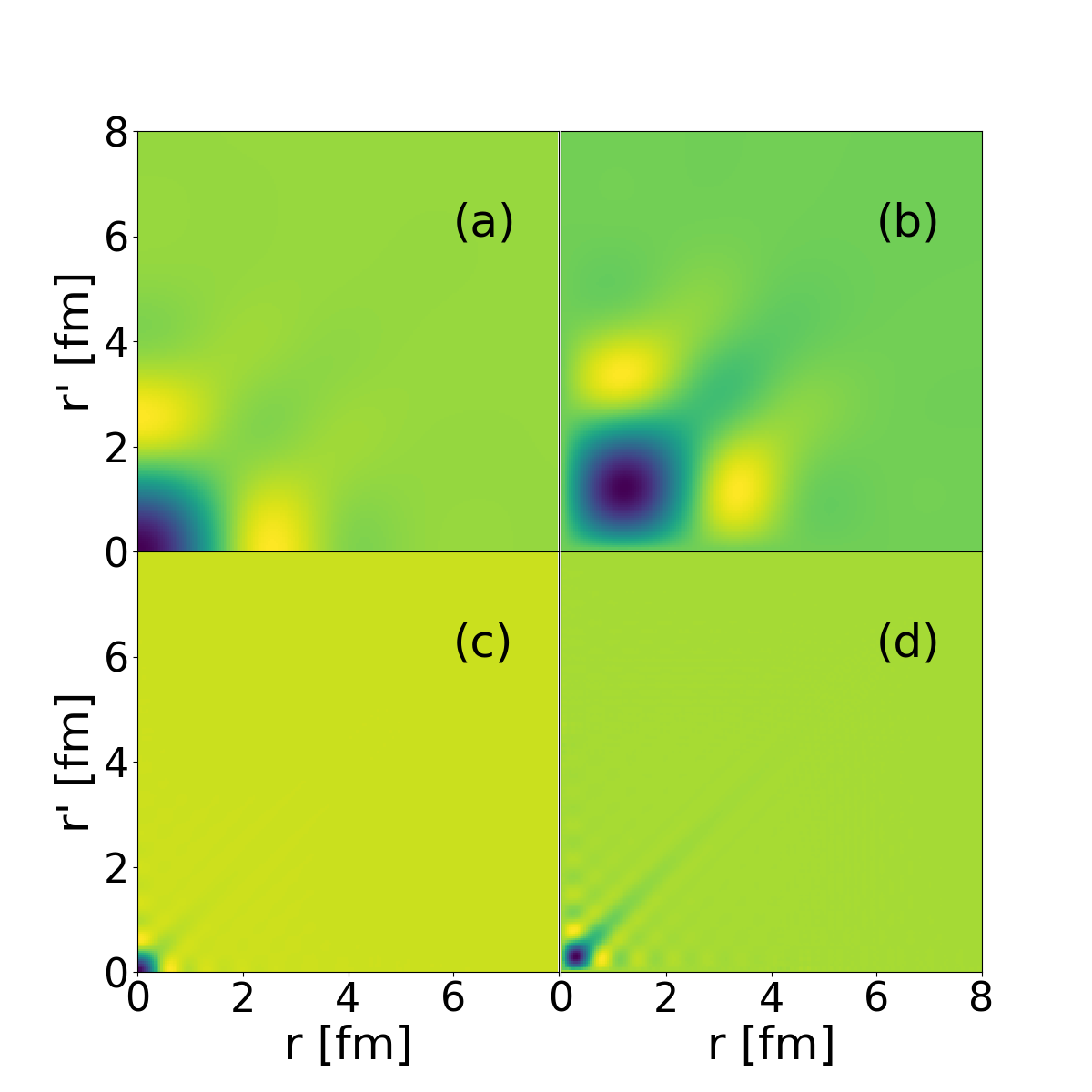}
\end{center}
\caption{[Color online] Radial dependence of the real part of the separable interaction obtained for  $^{48}$Ca(n,n) at 5 MeV:  (a) $J^{\pi}=1/2^+$ ($E_{range}=10$ MeV);  (b) $J^{\pi}=3/2^-$ ($E_{range}=10$ MeV); (c) $J^{\pi}=1/2^+$ ($E_{range}=2400$ MeV);  (d) $J^{\pi}=3/2^-$ ($E_{range}=2400$ MeV). The pale (yellow) and dark (blue) colors correspond to the  maxima and minima, respectively.
The color scale varies significantly between panels; in this figure we focus only the geometry of the potentials. }
 \label{fig-pot-ca5}
\end{figure}

\begin{figure}[h!]
\begin{center}
    \includegraphics[width=0.45 \textwidth]{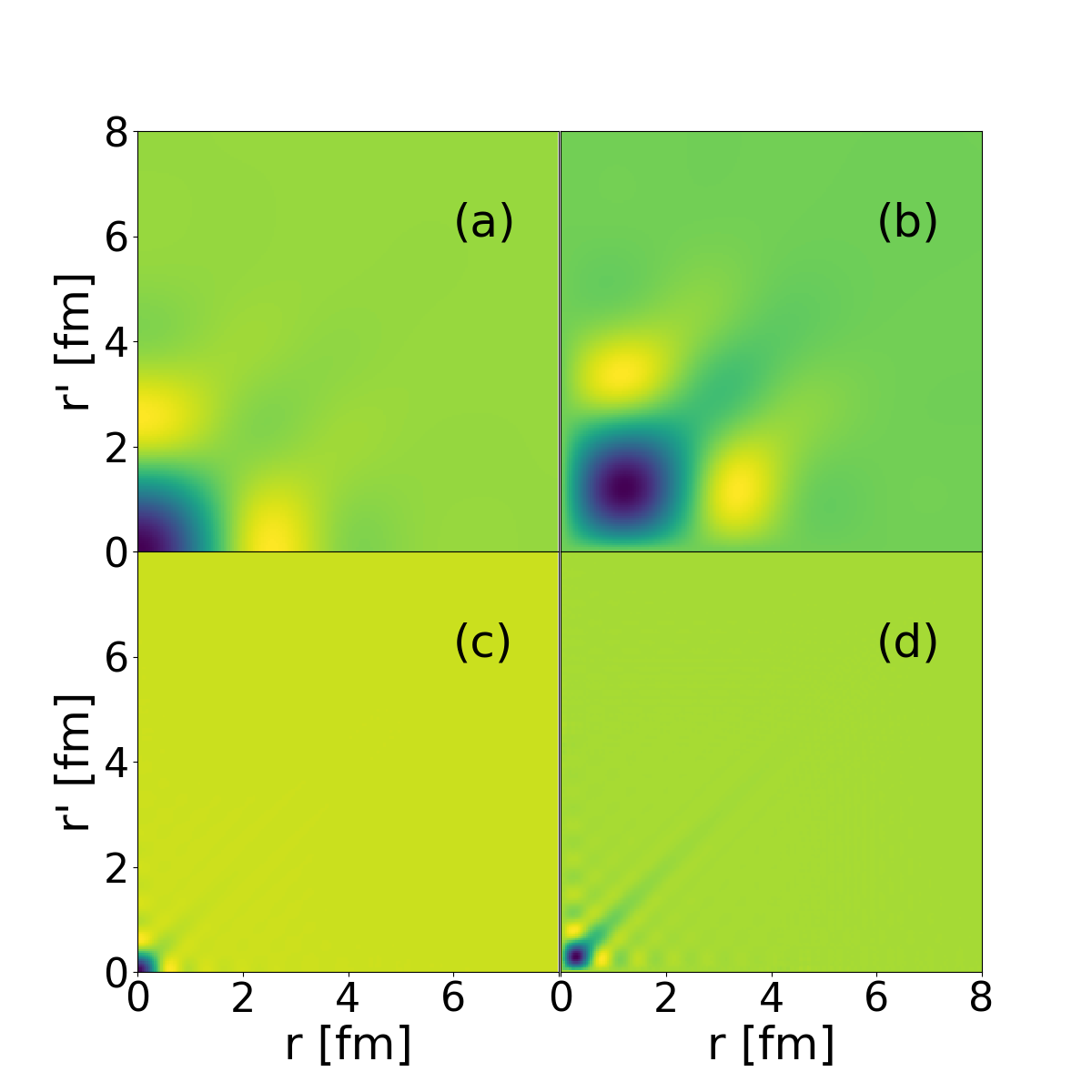}
\end{center}
\caption{[Color online] Radial dependence of the real part of the separable interaction obtained for  $^{48}$Ca(n,n) at 20 MeV: (a) $J^{\pi}=1/2^+$ ($E_{range}=20$ MeV); (b) $J^{\pi}=3/2^-$ ($E_{range}=20$ MeV); (c) $J^{\pi}=1/2^+$ ($E_{range}=2400$ MeV); (d) $J^{\pi}=3/2^-$ ($E_{range}=2400$ MeV). The pale (yellow) and dark (blue) colors correspond to the  maxima and minima, respectively. The color scale varies significantly between panels; in this figure we focus only the geometry of the potentials.}
 \label{fig-pot-ca20}
\end{figure}

Next we consider the radial dependence of the separable interactions. While the original CH89 optical potential  is local, the resulting separable interactions are non-local. To best illustrate this we present in Figs. \ref{fig-pot-ca5} and \ref{fig-pot-ca20} 
the real part of the n-$^{48}$Ca separable  potential  Re[$u_{\alpha}(r,r')$] for $E=$5 MeV and $E=$20 MeV respectively. We show two relevant partial waves ($J^{\pi}=1/2^+$ on the left panels and $J^{\pi}=3/2^-$ on the right panels) as well as the two extreme cases for the energy range (the lowest $E_{range}=10$ MeV on the top and the highest $E_{range}=2400$ MeV on the bottom). 


Several characteristics emerge from the analysis. 
\begin{itemize}
\item First of all, the separable interactions for $J^{\pi}=1/2^+; \ell=0$ have the minimum at $r=0$, independently of their rank or energy range. This is shown in panels (a) and (c) of  Figs.~\ref{fig-pot-ca5} and \ref{fig-pot-ca20}. Expectedly, introducing the repulsion from the centrifugal barrier shifts the minimum of the potential away from $r=0$ (shown in panels (b) and (d) for the partial wave $J^{\pi}=3/2^-; \ell=1$). 
\item Secondly, the potentials with smaller energy range (shown in panels (a) and (b)) have strong off-diagonal components. As we increase the energy range included in the EST procedure, the off-diagonal components shrink gradually toward the diagonal. Ultimately the potentials with the highest energy range  (shown in panels (c) and (d)) approach the diagonal form of the original CH89 potential. 
\item Thirdly, the off-diagonal structures of the separable interactions produced for E=5 MeV (Fig.~\ref{fig-pot-ca5}) are identical to those obtained for E=20 MeV (Fig.~\ref{fig-pot-ca20}). One should keep in mind that CH89 is energy dependent and therefore one might expect the corresponding separable interaction to be energy dependent too. We will return to this point in Section C.
\end{itemize}
The three broad features discussed before are persistent throughout our investigations, whether looking into the real or the imaginary parts of the potential, whether considering low or high angular momentum $\ell$. Also, the separable interactions generated for $^{16}$O($n$,$n$)$^{16}$O have the same qualitative characteristics as those shown in Figs.~\ref{fig-pot-ca5} and \ref{fig-pot-ca20}.

\begin{figure}[t!]
\begin{center}
\includegraphics[width=0.4 \textwidth]{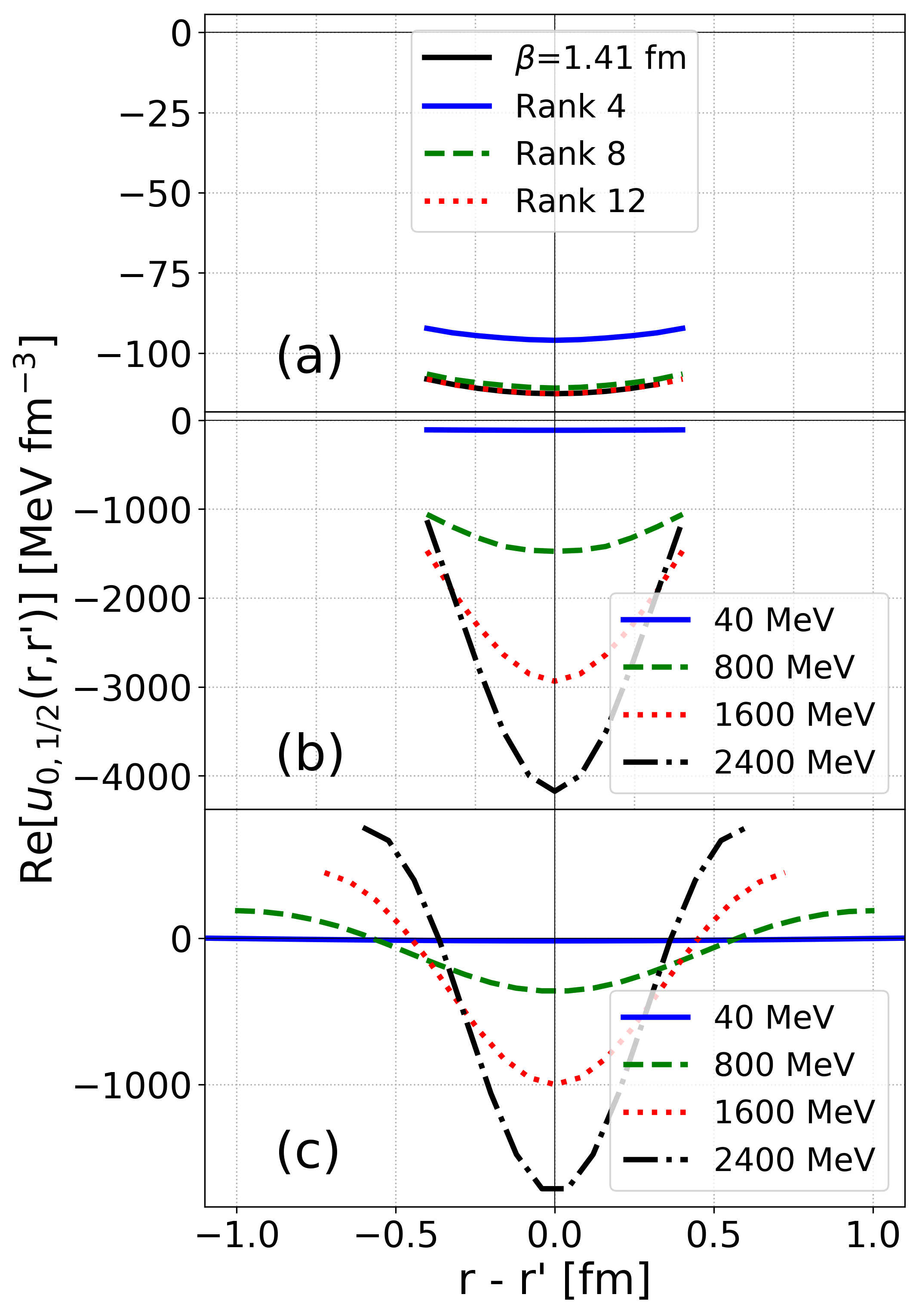}
\end{center}
\caption{[Color online] Cross diagonal of the real part of the separable interaction obtained for  $^{48}$Ca(n,n) at 5 MeV, $J^{\pi}=1/2^+$:  (a) comparing different ranks for $E_{range}=40$ MeV, with $(r+r')=0.4$ fm (b) comparing different $E_{range}$  with $(r+r')=0.4$ fm (c) comparing different $E_{range}$ through the potential maxima.}
 \label{fig-cross-l0}
\end{figure}

\begin{figure}[t!]
\begin{center}
\includegraphics[width=0.4 \textwidth]{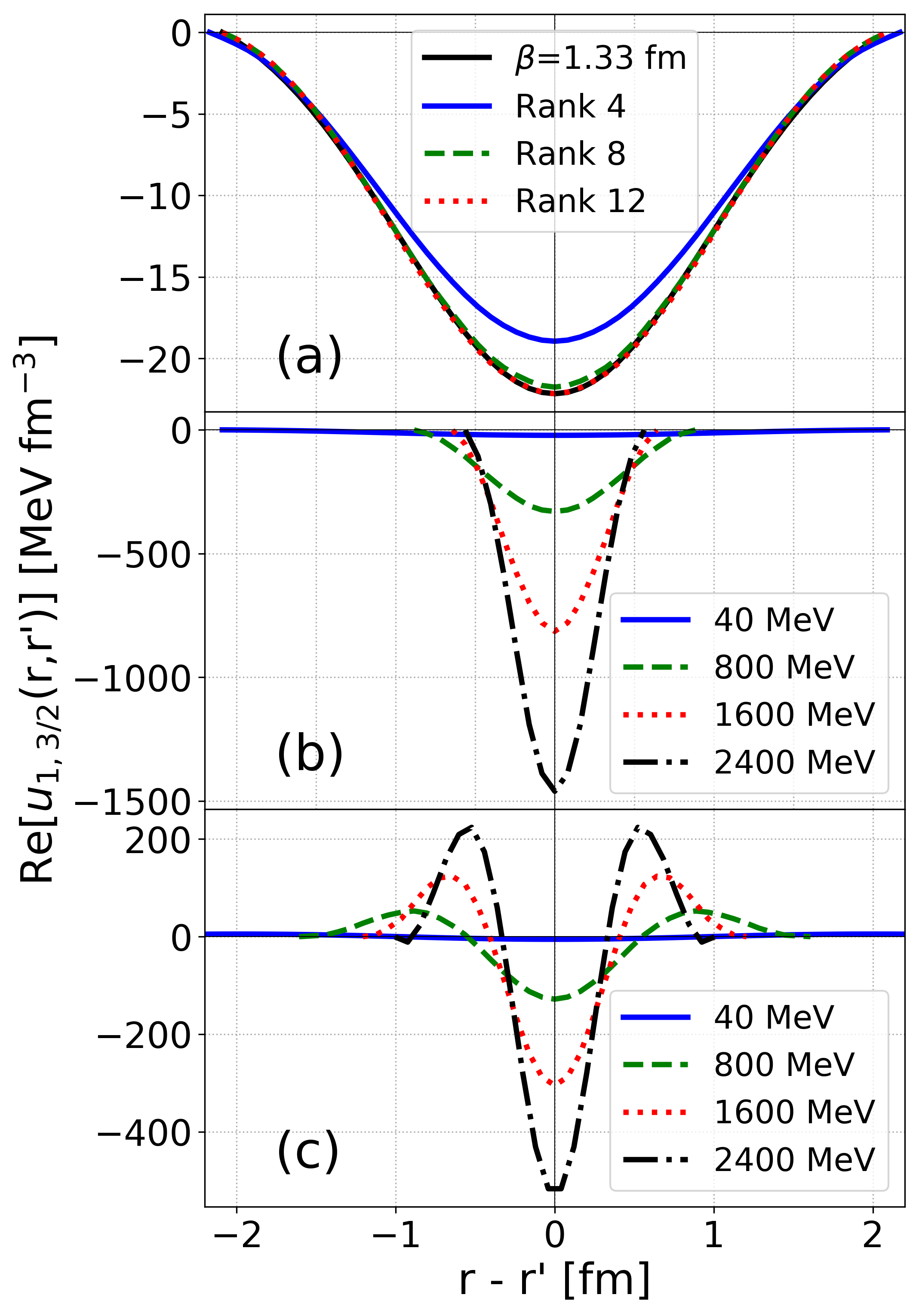}
\end{center}
\caption{[Color online] Cross diagonal of the real part of the separable interaction obtained for  $^{48}$Ca(n,n) at 5 MeV, $J^{\pi}=3/2^-$:  (a) comparing different ranks for $E_{range}=40$ MeV, through the potential minima (b) comparing different $E_{range}$ through the potential minima (c) comparing different $E_{range}$ through the potential maxima.}
 \label{fig-cross-l1}
\end{figure}

In order to best quantify the off-diagonal properties, we consider the separable potential cross-diagonals by plotting Re[$u_{\alpha}(r,r')$] as a function of $(r-r')$, while fixing $(r+r')$, in the region where the potential has its deep pocket. For $\ell=1$, we take $(r+r')$ such that the cross diagonal curve goes through the minimum of each potential. For $\ell=0$ these minima occur at the origin, so we instead fix $(r+r')$ at 0.4 fm.
In Fig. \ref{fig-cross-l0} we show the cross-diagonal behavior for the $\ell=0$ potential for neutrons on $^{48}$Ca at 5 MeV. A similar plot is shown in Fig. \ref{fig-cross-l1}  for $\ell=1$ neutrons on $^{48}$Ca also at 5 MeV. 

Figs.~\ref{fig-cross-l0}a and \ref{fig-cross-l1}a show the convergence of the potential with rank for a fixed energy range $E_{range}=40$ MeV. Results for $N=12$ are already converged and the behavior of the cross diagonal in this deep pocket is approximately Gaussian.
In contrast, Figs.~\ref{fig-cross-l0}b and \ref{fig-cross-l1}b show a very strong dependence of the cross diagonal potentials with the energy range included when calculating the separable interaction. With increasing $E_{range}$, the  interactions become deeper and more localized. In addition, we can analyze the cross-diagonal plots in the surface region, when the interaction reaches its maximum (Figs.~\ref{fig-cross-l0}c and \ref{fig-cross-l1}c). The cross-diagonal plots show a strong dependence on the $E_{range}$. This behavior merits further inspection.

\subsection{Non-locality behavior of the separable expansion}

In order to quantify the non-locality induced in the interaction we use the form introduced by Perey and Buck \cite{perey-buck}. Perey-Buck assume that the nonlocality of the optical potential is Gaussian:
\begin{equation}
\label{pb-eq}
    U^{PB}(\mathbf{r},\mathbf{r'}) = \exp \left( -\left| \frac{\mathbf{r} - \mathbf{r'}}{\beta}\right|^2  \right) U_{WS}\left(\frac{\mathbf{r} + \mathbf{r'}}{2}\right)
\end{equation}
where $U_{WS}$ is a local Woods-Saxon form and  $\beta$ is the  nonlocality parameter.

The partial-wave-decomposed interaction takes the form (See \cite{perey-buck}):
\begin{eqnarray}\nonumber
    u^{PB}_{\ell}(r,r') &=  \frac{2i^{\ell}}{\pi^{\frac{1}{2}}\beta} U_{WS}(\frac{1}{2}(r+r'))\\\label{eq:pb-pw-eq}
    & \times j_{\ell}(-i\frac{2rr'}{\beta^2})\exp\left(- \; \frac{r^2 + r'^2}{\beta^2}\right)\;.
\end{eqnarray}
In order to quantify the nonlocality,  Eq.(\ref{eq:pb-pw-eq}) was used to fit the cross-diagonal shapes of Figs.\ref{fig-cross-l0}a and \ref{fig-cross-l1}a and extract the nonlocality parameter $\beta$. In the fits, an arbitrary normalization was chosen and only the cross diagonal behavior at a fixed $r+r'$ was considered. These fits are shown by the black solid lines in Figs.\ref{fig-cross-l0}a and \ref{fig-cross-l1}a for $^{48}$Ca(n,n) at 5 MeV.


We repeat this procedure for each $E_{range}$ considered. 
and find that, consistently, around the minimum, and in the vicinity of $(r-r')=0$, the separable interaction can  be approximated by the functional form  in Eq.(\ref{eq:pb-pw-eq}). However, outside the deep pocket of the potential, the behavior is not well represented by the Perey and Buck form and, for that reason, we also study other measures of nonlocality. Particularly, we consider the distance between innermost roots $\Delta_{roots}$ and the distance between peaks $\Delta_{peaks}$  of the interaction, along the cross diagonal in the surface region depicted in Figs. \ref{fig-cross-l0}(c) and \ref{fig-cross-l1}(c).

\begin{figure}[t!]
\begin{center}
    \includegraphics[width=0.4 \textwidth]{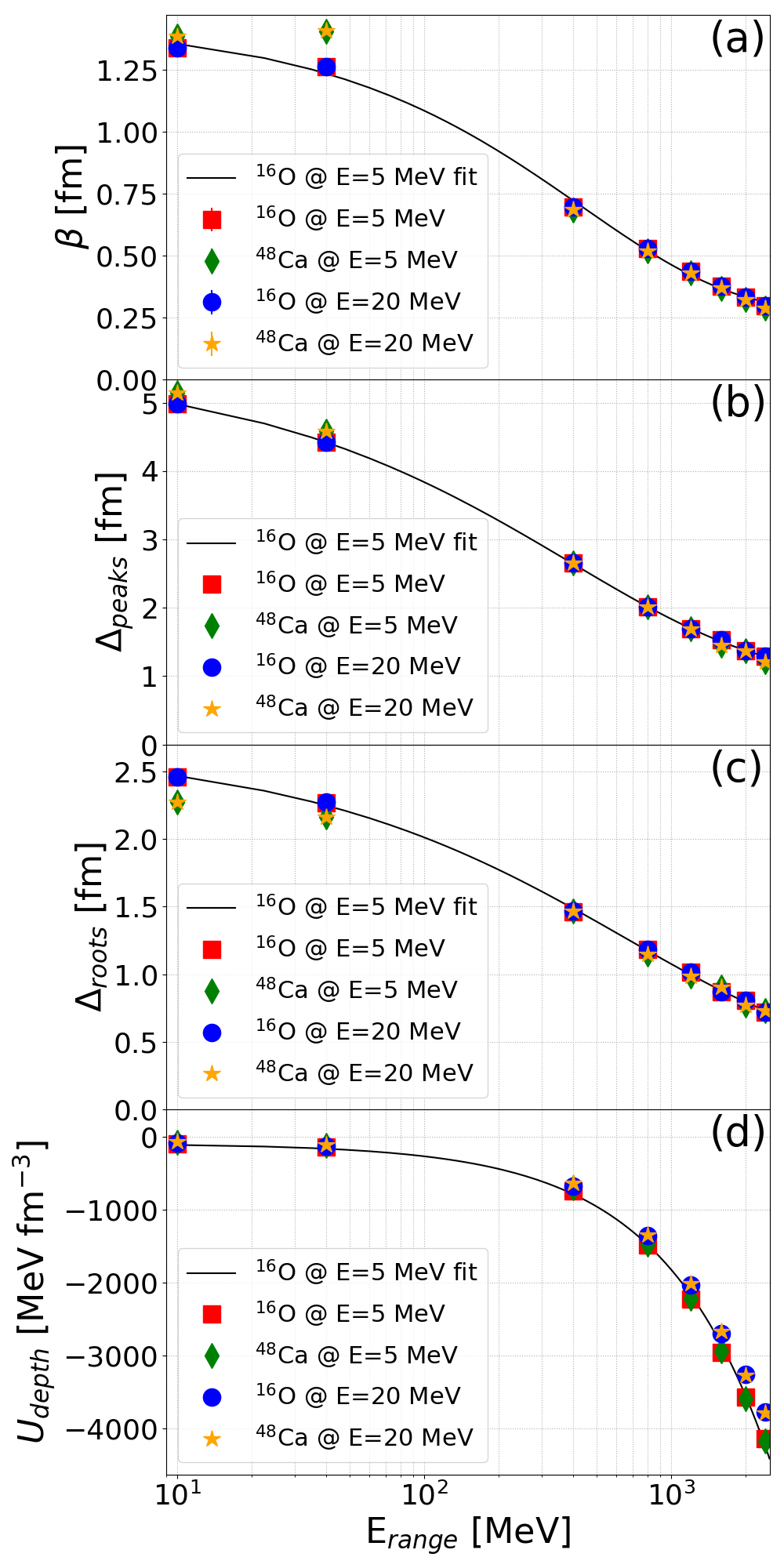}
\end{center}
\caption{ [Color online] Non-locality properties as a function of the energy range for  $\ell=0$ interactions: a) the non-locality parameter $\beta$; b) the distance between peaks $\Delta_{peaks}$ (fm); the distance between roots $\Delta_{roots}$ (fm); the depth in the minimum $U_{depth}$ (MeV).}
 \label{fig-beta0}
\end{figure}
\begin{figure}[t!]
\begin{center}
    \includegraphics[width=0.4 \textwidth]{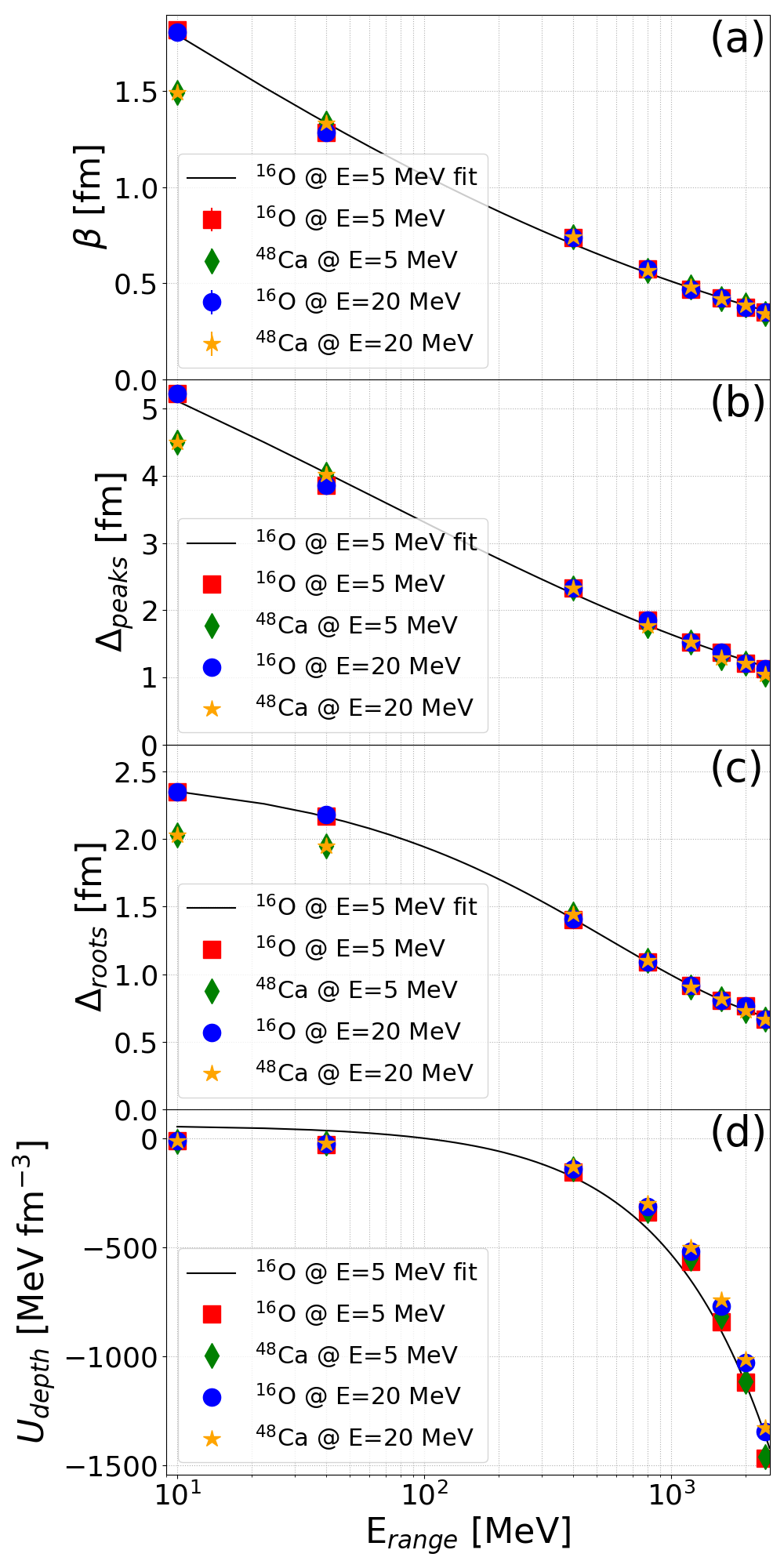}
\end{center}
\caption{[Color online] Non-locality properties as a function of the energy range for  $\ell=1$ interactions: a) the non-locality parameter $\beta$; b) the distance between peaks $\Delta_{peaks}$ (fm); the distance between roots $\Delta_{roots}$ (fm); the depth in the minimum $U_{depth}$ (MeV).}
 \label{fig-beta1}
\end{figure}

The results for $\beta$ as a function of $E_{range}$ are compiled in Figs. \ref{fig-beta0}(a) and  \ref{fig-beta1}(a) for $\ell=0$  and $\ell=1$.  We include all cases considered, namely neutrons on $^{16}$O at 5 MeV (red squares),  $^{48}$Ca at 5 MeV (green diamonds), $^{16}$O at 20 MeV (blue circles), $^{48}$Ca at 20 MeV (yellow stars). 
The values of $\beta$ include an error bar corresponding to one standard deviation obtained in the fit. 
In addition, Figs. \ref{fig-beta0}b and  \ref{fig-beta1}b show the distance between maxima along the cross diagonal 
$\Delta_{peaks}$ and  Figs. \ref{fig-beta0}c and  \ref{fig-beta1}c show the distance between the innermost roots of the potential $\Delta_{roots}$ along the cross diagonal taken through the maxima of the potential.
Finally,   the depths $U_{depth}$, defined as the minimum along the same cross-diagonal from which $\beta$ is also shown as a function of $E_{range}$ in Figs. \ref{fig-beta0}d and  \ref{fig-beta1}d. 

We first focus on the energy dependence of the separable potential. Figs. \ref{fig-beta0} and  \ref{fig-beta1} illustrate clearly that results for the nonlocal parameters for $E=5$ MeV are essentially identical to those for $E=20$ MeV (yellow stars are on top of green diamonds and blue circles are on top of red squares). As mentioned before, this may come as a surprise given that CH89 is strongly energy dependent. However, that energy dependence in the interaction is included through the model space  (namely $E_{range}$)  and becomes encoded in the off-diagonal $t$~matrix terms. 
It was shown in \cite{Feshbach:1958wf} (see page 18), that the optical model potential is intrinsically non-local and energy-dependent.  Imposing  locality on the optical potential introduces an additional form of energy dependence. 
The similarity of the results at 5~MeV and 20~MeV reveals that the intrinsic energy dependence the
optical model potential is weak, which is consistent with the findings of Ref.~\cite{Hlophe:2015rqn}. In that work it was observed that the $t$ matrices obtained with the energy-independent EST scheme~\cite{hlophe2013} were in close agreement with those obtained with the energy-dependent scheme discussed in Section~\ref{theory}
\footnote{To understand how this occurs, consider an energy-independent  non-local potential $\tilde U(r,r')$ so that $H=H_0+\tilde U$ has the eigenstates $\psi_E$, with $E$ being the energy eigenvalue. The local potential $U(r)$ that yields the same wavefunction fulfills  $U(r)\psi_E(r)=\int dr' r'^2 \tilde U(r,r')\psi_{E}(r')$.  Clearly $U(r)$ must be adjusted for each value of the energy in order to reproduce the wavefunctions corresponding $\tilde U(r,r')$, so that $U(r)\equiv U(r,E)$. When the EST scheme is invoked, it seeks to obtain a general non-local potential that reproduces the wavefunctions  $\psi_E$ across a whole energy range, although it does not contain the energy dependence induced by localization. }.

 
Second,  we examine the dependence of the nonlocal behavior of the  separable potential with the target. For  small values of the energy range, i.e., $E_{range}<$~50~MeV, the results for $^{16}$O differ from those corresponding
to $^{48}$Ca. This is not surprising since the CH89 potential depends on the mass of the target. 
However, this difference vanishes as the energy range approaches hundreds of MeV. In fact, for a given channel, the
nonlocality is determined exclusively by the energy range. This suggests a universal
correlation between the nonlocality and  energy range. To understand how this universality arises  we recall that
the basis states for the EST expansion are given by the Lippmann-Schwinger equation $|\psi_{E_i}\rangle= |p_i\rangle+G_0(E_i)U(E_i)|\psi_{E_i}\rangle$. The term containing the potential is thus inversely proportional to the energy, so that
$|\psi_{E_i}\rangle= |p_i\rangle$ in the limit $E_i\rightarrow\infty$.  Therefore, the potential should approach a limit
that is independent of the details of the original interaction for large energy ranges.

 Thirdly, although qualitatively the same, these properties are quantitatively dependent on the partial wave considered. In fact, when increasing the angular momentum, the deviation from the Gaussian form is more pronounced. However, as mentioned before, around the minimum, the Gaussian form is a good approximation.

While the non-locality parameter varies significantly within the energy ranges considered, one does expect it to go to zero when $E_{range} \rightarrow \infty$ because the original CH89 potential is local. To investigate this, we next consider the functional dependence of $\beta(E_{range})$ and  fit its $E_{range}$ dependence with to two trial functions, the first assuming the behavior is exponential and the second assuming the behavior is a power law:
\begin{eqnarray}
\beta_1 (E_{range}) &=& a \cdot \exp (b \cdot E_{range}^c) + d  \;, \\
\beta_2 (E_{range}) &=& \frac{a}{(E_{range} + b)^{c}} + d \;. \\
\end{eqnarray}
The results for the exponential fit of neutrons on $^{16}$O at 5 MeV are plotted in Figs. \ref{fig-beta0} and  \ref{fig-beta1} (black solid line). As expected, the results are mostly consistent with $\beta(E_{range} \rightarrow \infty) = 0$. However, the rate of convergence differs strongly with angular momentum and is always slower for the partial waves with higher angular momentum.
 For the other nonlocality measures, the fits (solid black lines in panels (b) and (c) of  Figs. \ref{fig-beta0} and  \ref{fig-beta1}) assume the same exponential form as for $\beta$, while for the depths (solid black lines in Figs. \ref{fig-beta0}(d) and  \ref{fig-beta1}(d)), the fit is linear. 
 
 The asymptotic value $\beta(E_{range} \rightarrow \infty) = d$ for all cases here considered are summarized in Table \ref{tab:beta}. Note that in all practical applications used before \cite{hlophe2017,lei2018}, the range considered was $E_{range}=50$ MeV, and therefore the potentials would exhibit strong non-locality.

\begin{table}[b]
\centering
\caption{Parameters for the fits of $\beta(E_{range})$.  }
\vspace{3mm}
\begin{tabular}{|c|c|c|r|r|}
\hline 
 E (MeV)  &	 Target &     Type           & $d$ ($J=1/2^+$)           &  $d$  ($J=3/2^-$) \\ \hline
5	& $^{16}$O	&        Exponential  &  0.27 $\pm$ 0.05  &  -0.03 $\pm$ 0.60\\
 	&	&       Power Law    &  -0.04 $\pm$ 0.23  &  -0.79 $\pm$ 0.85 \\
20	& $^{16}$O	&        Exponential  &  0.27 $\pm$ 0.05  &  -0.06 $\pm$ 0.62\\
 	&	&       Power Law    &  -0.04 $\pm$ 0.24  &  -0.86 $\pm$ 0.88 \\
\hline 
 5   & $^{48}$Ca 	&      Exponential  &  0.31 $\pm$ 0.06  &  0.29 $\pm$ 0.03\\
   	&          &       Power Law    &  -0.09 $\pm$ 0.82  &  -0.07 $\pm$ 0.06 \\
20 &   $^{48}$Ca &        Exponential  &  0.31 $\pm$ 0.06  &  0.29 $\pm$ 0.03\\
    &    	&    Power Law    &  -0.09 $\pm$ 0.83  &  -0.07 $\pm$ 0.06\\
\hline
\end{tabular}
\label{tab:beta}
\end{table}

\subsection{Connection with other frameworks}

Now that the properties of the nonlocal separable potentials have been uncovered, it is useful to compare the separable EST to other approaches. We first discuss the assumptions by Perey and Buck \cite{perey-buck}. Perey and Buck use a Gaussian form for the non-locality, estimate the nonlocality parameter to be $\beta=0.85$ fm and obtain the remaining parameters of the interaction from fitting angular distributions of neutrons scattering off $^{208}$Pb  at 7 MeV and 14.5 MeV. The separable interaction we obtain based on a global phenomenological potential for this energy range  is  $\beta=0.89-0.97$ fm for $\ell=0$, consistent with Perey and Buck's assumptions. The value for the nonlocal parameter is significantly larger for higher partial waves, $\beta=1.46-1.59$ fm for $\ell=1$. It is important to realize that the original Perey and Buck phenomenological interaction has no energy dependence nor target dependence except for the standard radius scaling with the mass).

Next we consider microscopic potentials such as \cite{rotureau2017,rotureau2018}. These are generated from a truncated many-body framework which effectively impose an $E_{range}$ in the calculation of the optical potential. Although the behavior of the microscopic optical potential is not Gaussian, the overall shape of $u(r,r')$ cross-diagonals for the case of $^{48}$Ca \cite{rotureau2018} are similar to those shown in Fig. \ref{fig-cross-l1}.  The microscopic interaction exhibits  $\hat\beta\approx 1$ fm. For $^{16}$O \cite{rotureau2017}, the cross diagonals are sufficiently different that a quantitative comparison makes no sense. If we use our separable framework to determine the effective energy range included in a given interaction, one would conclude that the microscopic optical potentials of \cite{rotureau2017,rotureau2018} contain physics in the region $E_{range}<10$ MeV.


\section{Conclusions}
\label{conclusions}

Since the EST separable method is now being applied to nucleon-nucleus optical potentials for nuclear reactions calculations \cite{hlophe2013,hlophe2014,hlophe2017}, it is important to understand in detail the properties this procedure is inducing in the interactions. Of particular importance is the nonlocality, which has been shown to modify reaction observables. With this goal in mind, we have performed a systematic study, for neutron scattering on two stable targets ($^{16}$O and $^{48}$Ca) at two beam energies $E=5$ MeV and $E=20$ MeV. Starting from a local phenomenological optical potential, we have generated separable interactions that represent the neutron scattering process. We have studied the convergence with the rank  and the energy range included in constructing the interaction. 

We find that the separable procedure induces a large nonlocality in the interaction when $E_{range}<50$ MeV.  Morever, we observe that, even when including in the expansion many support points with energy ranges up to $0\le E\le 2400$~MeV, the resulting potential retains non-local behavior.
This nonlocality becomes considerably smaller as $E_{range}$ is increased, eventually tending to zero as $E_{range} \rightarrow \infty$ as expected. While for small $E_{range}$ the magnitude of the nonlocality depends on the target, this dependence is washed away for increasing $E_{range}$, following a universal curve. 
Focusing on the deep pocket of the separable interaction, for all cases we find that the  nonlocality increases with angular momentum.

While there is a strong dependence of the separable optical potential with $E_{range}$ used to construct it, there is virtually no dependence on the beam energy. The strong energy dependence in the original phenomenological optical potential  disappears once nonlocality is allowed in the interaction.

We also compare our results with other studies on nonlocal optical potentials. We find that overall the separable interactions are not well described by the Gaussian form used by Perey and Buck \cite{perey-buck}. However, around the minimum, they can be approximated by a Gaussian form and for s-waves the magnitude of nonlocality we obtain is similar to that assumed by \cite{perey-buck}. We also compare our interactions for $^{48}$Ca with those obtained from ab-initio calculations \cite{rotureau2018}. 

In closing, it is useful to think about the EST procedure in the context of renormalization group theory. In \cite{bogner2003plb}, the effective potential is defined through the Block-Horowitz equation. This equation explicitly re-sums all the higher momentum modes while preserving the low-energy momentum scattering amplitudes. Because the effective interaction produced this way is energy dependent,  another transformation is needed to arrive at an interaction that is only momentum dependent $V_{low k}(k,k')$. 
 A direct comparison between EST and $V_{low k}$ is currently not possible (because $V_{low k}$ has not been applied to optical potentials) but could be very enlightenning. For both the EST and $V_{low k}$ schemes, the potentials are constrained so that the bound and scattering states of the original interaction are reproduced over a finite energy range. As such, both techniques have the effect of shuffling high momentum components into non-local behavior of the effective interaction.

\begin{acknowledgments}
We thank Scott Bogner and Charlotte Elster for useful discussions. This work was supported by the National Science Foundation under Grant  PHY-1811815.  This work relied on iCER and the High Performance Computing Center at Michigan State University for computational resources. 
\end{acknowledgments}

\bibliography{separable-v6} 

\end{document}